\documentclass[
    superscriptaddress,
    reprint,
    showkeys,
    showpacs,
    amsmath,
    amssymb,
    aps,
    prb,
    twocolumn,
    floatfix
]{revtex4-1}
\usepackage{amsmath}
\usepackage{physics}
\usepackage{siunitx}
\usepackage{graphicx}
\usepackage{dcolumn}
\usepackage{bm}
\usepackage{braket}
\usepackage[caption=false]{subfig}
\usepackage{float}
\usepackage{booktabs}
\usepackage{multirow}
\usepackage{tabularx}
\usepackage{color}
\definecolor{link}{RGB}{57,106,177}
\definecolor{blue}{RGB}{0,0,0}
\newcommand{\ped}[1]{\ensuremath{_{\rm #1}}}
\newcommand{\apex}[1]{\ensuremath{^{\rm #1}}}

\newcommand{\mywidth}{0.95\columnwidth}

\usepackage{hyperref}
\hypersetup{
    colorlinks=true,
    linkcolor=link,
    citecolor=link,
    urlcolor=link
}
\AtBeginDocument{
\heavyrulewidth=.08em
\lightrulewidth=.05em
\cmidrulewidth=.03em
\belowrulesep=.65ex
\belowbottomsep=0pt
\aboverulesep=.4ex
\abovetopsep=0pt
\cmidrulesep=\doublerulesep
\cmidrulekern=.5em
\defaultaddspace=.5em
}
\begin{document}
\title{Strong band-filling-dependence of the scattering lifetime in gated MoS\ped{2} nanolayers induced by the opening of intervalley scattering channels}
\author{Davide Romanin}
\affiliation{\mbox{
Department of Applied Science and Technology, Politecnico di Torino, I-10129 Torino, Italy}
}
\author{Thomas Brumme}
\affiliation{\mbox{
Wilhelm-Ostwald-Institut f{\"u}r Physikalische und Theoretische Chemie, Linn{\'e}stra{\ss}e 2, 04103 Leipzig, Germany}
}
\affiliation{\mbox{
Theoretische Chemie, Technische Universit{\"a}t Dresden, Bergstra{\ss}e 66c, 01062 Dresden, Germany}
}
\author{Dario Daghero}
\affiliation{\mbox{
Department of Applied Science and Technology, Politecnico di Torino, I-10129 Torino, Italy}
}
\author{Renato S. Gonnelli}
\email{renato.gonnelli@polito.it}
\affiliation{\mbox{
Department of Applied Science and Technology, Politecnico di Torino, I-10129 Torino, Italy}
}
\author{Erik Piatti}
\email{erik.piatti@polito.it}
\affiliation{\mbox{
Department of Applied Science and Technology, Politecnico di Torino, I-10129 Torino, Italy}
}

\begin{abstract}
Gated molybdenum disulphide (MoS$_2$) exhibits a rich phase diagram upon increasing electron doping, including a superconducting phase, a polaronic reconstruction of the bandstructure, and structural transitions away from the 2H polytype. The average time between two charge-carrier scattering events -- the scattering lifetime -- is a key parameter to describe charge transport and obtain physical insight in the behavior of such a complex system. In this work, we combine the solution of the Boltzmann transport equation (based on \textit{ab-initio} density functional theory calculations of the electronic bandstructure) with the experimental results concerning the charge-carrier mobility, in order to determine the scattering lifetime in gated MoS$_2$ nanolayers as a function of electron doping and temperature. From these dependencies, we assess the major sources of charge-carrier scattering upon increasing band filling, and discover two narrow ranges of electron doping where the scattering lifetime is strongly suppressed. We indentify the opening of additional intervalley scattering channels connecting the simultaneously-filled K/K\apex{\prime} and Q/Q\apex{\prime} valleys in the Brillouin zone as the source of these reductions, which are triggered by the two Lifshitz transitions induced by the filling of the high-energy Q/Q\apex{\prime} valleys upon increasing electron doping. 
\end{abstract}


\maketitle

\section{\label{sec:intro}Introduction}

In the last decade, the ionic gating technique has become a fundamental tool for probing the ground-state properties of low-dimensional systems as a function of doping. Indeed, thanks to the field-effect transistor (FET) architecture it is possible to investigate the rich phase diagrams of (quasi) two-dimensional (2D) materials and surfaces in an almost continuous way~\cite{YeNatMater2010, YeScience2012, JoNanoLett2015, ShiSciRep2015, YuNatNano2015, SaitoACSNano2015, PiattiJSNM2016, LiNature2016, WangNature2016, XiPRL2016, OvchinnikovNatCommun2016, ShiogaiNatPhys2016, LeiPRL2016, PiattiPRB2017, ZengNanoLett2018, DengNature2018, WangNatNano2018, PiattiPRM2019, PiattiApSuSc2020, RenNL2019}. The transition metal dichalcogenides (TMDs) represent a notably tunable class of materials thanks to the occurrence of both superconducting (SC) and charge-density-wave (CDW) phases~\cite{KlemmBook2012, KlemmPhysC2015}. Among them, molybdenum disulphide ($2H$-MoS\ped{2}) has been the most studied both theoretically and experimentally, {\color{blue}owing to its stability at ambient pressure and temperature, the ease by which it can be exfoliated, its sizeable bandgap~\cite{WangNatNano2012} and the indirect-to-direct gap transition that it undergoes when thinned from the bulk to the single-layer~\cite{WangNatNano2012, MakPRL2010, SplendianiNL2010}, which make it eminently suitable for electronic and optoelectronic applications~\cite{WangNatNano2012, FerrariNanoscale2015, MakNatPhotonics2016}.} This layered semiconductor develops a SC phase with a maximum transition temperature $T_c \sim 11$~K either via ion intercalation~\cite{ZhangNanoLett2016, PiattiAPL2017} or by electrostatic ion accumulation at the interface between the material and an electrolyte~\cite{YeScience2012, BiscarasNatCommun2015, CostanzoNatNano2018}.

\begin{figure}[]
\centering
\includegraphics[width=1.0\columnwidth]{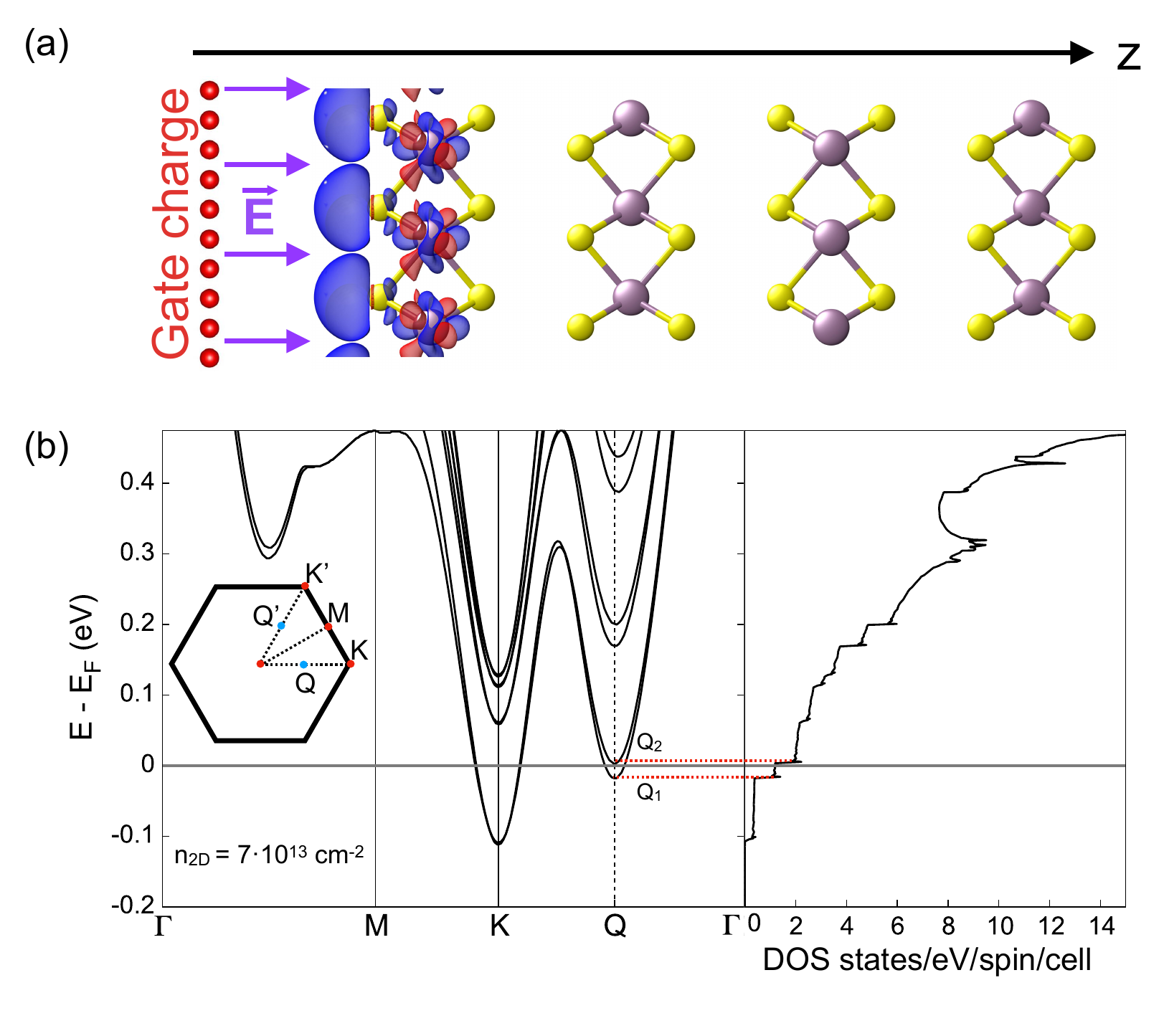}
\caption{
(a) Schematic view of the 4L-MoS\ped{2} crystal in the FET configuration. Yellow spheres are S atoms and purple-grey spheres are Mo atoms. The positive ions accumulated at the electrolyte-sample interface are represented by red spheres. The negative (positive) induced charge at the surface of the 4L-MoS\ped{2} crystal is depicted with blue (red) clouds around atoms {\color{blue}using isosurfaces at $1/15$th of the maximum charge density}. The gate electric field $\vec{E}$ is depicted as violet arrows.
(b) Electronic bandstructure and density of states (DOS) of gated 4L-MoS\ped{2} at an electron doping level $n\ped{2D}=7\times10^{13}$~cm\apex{-2}. The grey line represents the Fermi energy $E\ped{F}$. The inset shows the First Brillouin Zone of 2$H$-MoS\ped{2} where the $\Gamma$, K/K\apex{\prime} and M high-symmetry points are highlighted. Band edges Q\ped{1} and Q\ped{2} of the spin-orbit split sub-bands at the Q/Q\apex{\prime} point are highlighted by red-dashed lines.}
\label{fig:structure}
\end{figure}

When $2H$-MoS\ped{2} is electrostatically electron-doped in the FET configuration (Fig.~\ref{fig:structure}a), the presence of the electric field along the direction orthogonal to the surface breaks inversion symmetry and leads to a Zeeman-like spin-orbit splitting of the conduction bands~\cite{KormanyosPRB2013, YuanPRL2014} in the Brillouin Zone (BZ). The conduction band minima lie at the inequivalent K/K\apex{\prime} points (located at the corner of the hexagonal BZ) and Q/Q\apex{\prime} points (which lie more or less half-way between K/K\apex{\prime} and the center of the BZ $\Gamma$), as depicted in the inset of Fig.~\ref{fig:structure}b. The corresponding spin-split electron pockets are the so-called valleys common to all TMDs in the $2H$ crystal structure~\cite{BrummePRB2015,  BrummePRB2016, KangNanoLett2017, RoldanAnnPhys2014, ZhaoACR2015} that become filled upon electron doping. As a consequence, the geometry of the Fermi surface (FS) of gated MoS\ped{2} strongly depends on their occupation. Such valley filling is in turn strongly dependent on the number of layers, on the strength of the electric field, and on the tensile strain of the sample~\cite{BrummePRB2015, BrummePRB2016}. {\color{blue}The Zeeman-like spin-orbit splitting is crucial in determining the properties of the gate-induced SC state~\cite{YuanPRL2014}, as it leads to the spin-valley locking of the Copper pairs~\cite{LuScience2015, SaitoNatPhys2016} and the so-called 2D Ising SC and its ultrahigh out-of-plane critical magnetic field~\cite{LuScience2015, SaitoNatPhys2016}.}

The filling of the K/K\apex{\prime} and Q/Q\apex{\prime} valleys can be probed experimentally by means of electric transport measurements. When higher-energy sub-bands are crossed by the Fermi level, characteristic kinks appear in the doping-dependence of the conductivity of ion-gated TMD nanolayers~\cite{PiattiNL2018,ZhangNL2019}. This also allows directly probing the change in the topology of the Fermi surface, i.e. the occurrence of Lifshitz transitions: At low doping only the K/K\apex{\prime} valleys are filled, giving rise to two electron pockets only; whereas as the electron doping increases the Q/Q\apex{\prime} valleys become filled as well, generating six new electron pockets~\cite{PiattiNL2018, BrummePRB2015, BrummePRB2016}.

Recent developments in density functional theory (DFT) allow computing the electronic~\cite{BrummePRB2015, BrummePRB2016, BrummePRB2014} and vibrational~\cite{SohierPRB2017} properties of materials in the FET configuration from first principles by fully taking into account the presence of an orthogonal electric field in a self-consistent way. In such a way, it has been possible to obtain the electronic structure of many gated TMDs~\cite{BrummePRB2015, BrummePRB2016}, to study the flexural phonons in graphene~\cite{SohierPRB2017}, to explore the anomalous screening of an electric field at the surface of niobium nitride~\cite{PiattiApSuSc2018nbn} and to predict a possible high-$T_c$ SC phase transition in diamond thin films~\cite{RomaninAPSUSC2019, RomaninApSuSc2020}.

More specifically, in Ref.~\onlinecite{PiattiJPCM2019} we showed that DFT calculations can reliably reproduce the experimental doping dependence both of the conductivity and of the valley filling in ion-gated MoS\ped{2} nanolayers, when the presence of the transverse electric field, the number of layers and the level of strain in the experimental samples are taken into account. However, our analysis provided no information on the charge-carrier scattering lifetime, which is a crucial physical quantity necessary to describe charge transport in the system. The scattering lifetime $\tau$ is the average time between two successive scattering events experienced by a given charge carrier, and it directly determines key parameters for both the physics of the system and the device operation, such as for example the charge-carrier mobility, the mean free path and the degree of metallicity of the system. In this work, we tackle this issue directly by following the approach introduced in Ref.~\onlinecite{BrummePRB2016}: We start by computing the \textit{ab initio} bandstructure of gated 4L-MoS\ped{2}, and subsequently combine the Hall mobility-to-lifetime ratio, obtained by solving the Boltzmann transport equation~\cite{BoltzTrap}, with the Hall mobility calculated from the doping-dependence of the conductivity reported in Refs.~\onlinecite{PiattiNL2018, PiattiJPCM2019}. We find that when the Q/Q\apex{\prime} valleys are filled by the increasing field-induced electron doping, i.e. when the Lifshitz transitions occur, the scattering lifetime undergoes a strong reduction. We show that this observation can in turn be linked to the opening of new intervalley scattering channels between the simultaneously-filled K/K\apex{\prime} and Q/Q\apex{\prime} electron pockets. We discuss how this phenomenon can strongly affect key properties of gated MoS\ped{2} reported in the literature, such as the electron-phonon coupling, the gate-induced SC state, the polaronic reconstruction of the K/K\apex{\prime} Fermi sea, and the low-temperature incipient localization often observed in real devices.

\section{\label{sec:methods}Methods}

\subsection{Computational details}

In order to precisely match the experimental conditions of Ref.~\onlinecite{PiattiNL2018}, we considered a four-layer MoS\ped{2} crystal (4L-MoS\ped{2}), set the in-plane lattice parameter to the experimental bulk value, and added $0.13\%$ tensile strain~\cite{PiattiJPCM2019}. We then performed the DFT calculations using the plane-wave pseudopotential method as implemented in Quantum ESPRESSO~\cite{QE, QE_2}. We made use of fully-relativistic projector-augmented pseudopotentials~\cite{BlochlPRB1994} and of the Perdew-Burke-Ernzerhof exchange-correlation functional~\cite{PerdewPRL1996} including van der Waals dispersion corrections~\cite{GrimmeJPC2006}. We set the energy cutoff for the wave functions to 50 Ry and that for the charge density to 410 Ry. We performed the Brillouin zone integration using a Monkhorst-Pack grid~\cite{MonkhorstPRB1976} of $32\times32\times1$ \textbf{k} points with a Gaussian broadening of 2 mRy, and set the self-consistency conditions for the solution of the Kohn-Sham equations to $10^{-9}$~Ry for the total energy, and to $10^{-4}$~Ry/Bohr for the total force acting on the atoms during the structure relaxation. After convergence of the ground-state density, we then performed an additional non-self-consistent calculation on a denser grid of $64\times64\times1$ \textbf{k} points that will be used later for accurate solution of the Boltzmann equation.

We modelled the FET setup using the method described in Refs.~\onlinecite{BrummePRB2014, BrummePRB2015, PiattiApSuSc2018nbn}, where a dipole correction is employed in order to get rid of spurious Coulomb interactions along the non-periodic direction due to repeated images of the system under study. We placed the dipole for the dipole correction at $z\ped{dip} = d\ped{dip}/2$ with $d\ped{dip}=0.01L$, $L$ being the size of the unit cell in the $z$-direction, and the charged plane mimicking the gate electrode slightly closer to the MoS\ped{2} crystal at $z\ped{mono} = 0.011L$. A potential barrier of height $V_0 = 2$~Ry and width $d\ped{b}=0.1L$ is placed between the gate and MoS\ped{2} crystal in order to prevent charge spilling. 

The Boltzmann transport equation is solved in the constant-relaxation-time approximation, i.e. $\tau_{i,\vb{k}}=\tau(E_F)$, as implemented in the BoltzTraP~\cite{BoltzTrap} code starting from the eigenvalues of the Kohn-Sham hamiltonian obtained after the non-self-consistent computation. The ratio between the number of planewaves and the number of band energies is set to 5. In order to solve the integrals for the computation of transport tensors, we took into account bands that fall into an energy window of $0.04$ Ry around the Fermi energy.

\subsection{Solution of the Boltzmann equation}

For 2D systems, the conductivity tensors $\sigma\ped{\alpha\beta}$ and  $\sigma\ped{\alpha\beta\gamma}$ at temperature $T$ and chemical potential $E\ped{F}$ are~\cite{BrummePRB2015, BrummePRB2016}:
\begin{equation}
\begin{split}
\sigma\ped{\alpha\beta}(T,E\ped{F}) = \frac{e^2}{4\pi^2}\sum_i\int\tau_{i,\textbf{k}}v\ped{\alpha}^{i,\textbf{k}}v\ped{\beta}^{i,\textbf{k}}\left[-\frac{\partial f_{E\ped{F}}(T,\varepsilon_{i,\textbf{k}})}{\partial \varepsilon}\right]d\textbf{k}
\end{split}
\label{eq:cond_BTE}
\end{equation}
\begin{equation}
\begin{split}
\sigma\ped{\alpha\beta\gamma}(T,E\ped{F}) = \frac{e^3}{4\pi^2}\sum_i\int\tau^2_{i,\textbf{k}}\epsilon\ped{\gamma \delta \rho}v\ped{\alpha}^{i,\textbf{k}}v\ped{\rho}^{i,\textbf{k}}\bigl(M\ped{\beta \delta}^{i,\textbf{k}}\bigr)^{-1}\\\times\left[-\frac{\partial f_{E\ped{F}}(T,\varepsilon_{i,\textbf{k}})}{\partial \varepsilon}\right]d\textbf{k}
\end{split}
\label{eq:cond_BTE2}
\end{equation}
where $e$ is the elementary charge, $\hbar$ is the reduced Planck constant, $\varepsilon_{i,\textbf{k}}$ is the energy of the $i$-th band at momentum $\textbf{k}=(k\ped{x},k\ped{y})$, $\epsilon\ped{\alpha \beta \gamma}$ is the Levi-Civita symbol, $v\ped{\alpha}^{i,\textbf{k}}=\hbar^{-1}\partial \varepsilon_{i,\textbf{k}} / \partial k\ped{\alpha}$ is the group velocity along the $\alpha$-th k-component, $\bigl(M\ped{\alpha \beta }^{i,\textbf{k}}\bigr)^{-1}=\hbar^{-2}\partial^2\varepsilon_{i,\textbf{k}} / \partial k\ped{\alpha}\partial k\ped{\beta}$ is the inverse mass tensor for the $\alpha$-th and $\beta$-th k-components and $f_{E\ped{F}}(T,\varepsilon)$ is the Fermi distribution function. Notice that for a general 3D system all of the indices $\{\alpha,\beta,\gamma,\delta,\rho\}$ are  run over all of the cartesian coordinates $\{x,y,z\}$, however in a 2D system $\gamma=z$ and $\{\alpha,\beta,\delta,\rho\}$ are limited to the in-plane coordinates $\{x,y\}$.

Thanks to the conductivity tensors it is possible to compute the Hall tensor as:
\begin{equation}
R\ped{ijk}=\bigl(\sigma\ped{\alpha j}\bigr)^{-1}\sigma\ped{\alpha\beta k}\bigl(\sigma\ped{i\beta}\bigr)^{-1}
\end{equation}

While the relaxation time $\tau_{i,\textbf{k}}$ can be both band- and momentum-dependent, in the often-used \textit{constant}-relaxation-time approximation one sets $\tau_{i,\textbf{k}} = \tau(E\ped{F})=\tau$ (where $E_F$ is the Fermi level). In this approximation, both $\sigma\ped{\alpha\beta}/\tau$ and the Hall tensor are independent of $\tau$ and can be directly computed with BoltzTraP~\cite{BoltzTrap} from the \textit{ab-initio} bandstructure obtained on a fine $k$ mesh. Thus, the theoretical value of the in-plane Hall mobility-to-lifetime ratio can be computed simply as~\cite{BrummePRB2016}:
\begin{equation}
\frac{\mu\ped{H}\apex{th}}{\tau}(T,E\ped{F}) = \frac{\sigma_{xx}}{\tau} (T,E\ped{F})\,\left[-R_{xyz}(T,E\ped{F})\right]
\label{eq:mu_tau}
\end{equation}
where $\sigma_{xx}/\tau$ is the in-plane conductivity-to-lifetime ratio and $R_{xyz}$ is the Hall coefficient, i.e. the component of the Hall tensor with the induced electric field along $y$, the current flowing along $x$, and the magnetic field applied along $z$ (see Fig.~\ref{fig:structure}~(b)). Here, we have made use of the fact that the conductivity tensor $\sigma_{\alpha\beta}$ of crystals with hexagonal symmetry, such as MoS\ped{2}, has only two independent components (in-plane $\sigma_{xx}$ and out-of-plane $\sigma_{zz}$)~\cite{AshcroftBook, BrummePRB2015, BrummePRB2016}. To allow for a more reliable comparison with the experimental results, from $R_{xyz}$ we also directly determine the Hall carrier density $n\ped{H} = -1/eR_{xyz}$, since in principle in TMDs $n\ped{H}$ is known to sometimes strongly deviate from the actual doping charge $n\ped{2D}$~\cite{BrummePRB2015, BrummePRB2016}.

Finally, we also compute the thermally-smeared 2D density of states (DOS) as:
\begin{equation}
\begin{split}
\text{DOS}(T,E) = \frac{e^2}{4\pi^2}\sum_i\int\left[-\frac{\partial f_{E\ped{F}}(T,\varepsilon_{i,\textbf{k}})}{\partial \varepsilon}\right]d\textbf{k}
\end{split}
\label{eq:smeared_DOS}
\end{equation}

\subsection{Determination of the scattering lifetime}

We determine the scattering lifetime by means of the approach originally developed in Ref.~\onlinecite{BrummePRB2016}, where it was applied to gated WS\ped{2}, and that has later been successfully applied to other gated materials such as few-layer graphene~\cite{PiattiApSuSc2017, Gonnelli2DMater2017} and epitaxial diamond films~\cite{PiattiApSuSc2020}. Specifically, once the dependence of $\mu\ped{H}\apex{th}/\tau$ as a function of $n\ped{H}$ is known, the scattering lifetime $\tau$ in a gated device can be easily obtained from the experimental values of the Hall mobility. Here, we directly calculate it as:
\begin{equation}
\mu\ped{H}\apex{exp} = \frac{\sigma_{xx}}{e~n\ped{H}}
\end{equation}
from the values of $\sigma_{xx}$ and $n\ped{H}$ we experimentally measured in Ref.~\onlinecite{PiattiNL2018} and summarized in Ref.~\onlinecite{PiattiJPCM2019}.
The scattering lifetime can then be recovered by:
\begin{equation}
\tau = \frac{\mu\ped{H}\apex{exp}} {\mu\ped{H}\apex{th}/\tau}
\label{eq:tau_from_mu}
\end{equation}
for any value of $n\ped{H}$ and $T$ for which both the experimental Hall mobility and the theoretical mobility-to-lifetime ratio have been determined.

\section{\label{sec:results}Results and discussion}
\subsection{\color{blue}Density of states and transport coefficients}
\begin{figure}[]
\centering
\includegraphics[width=0.9\columnwidth]{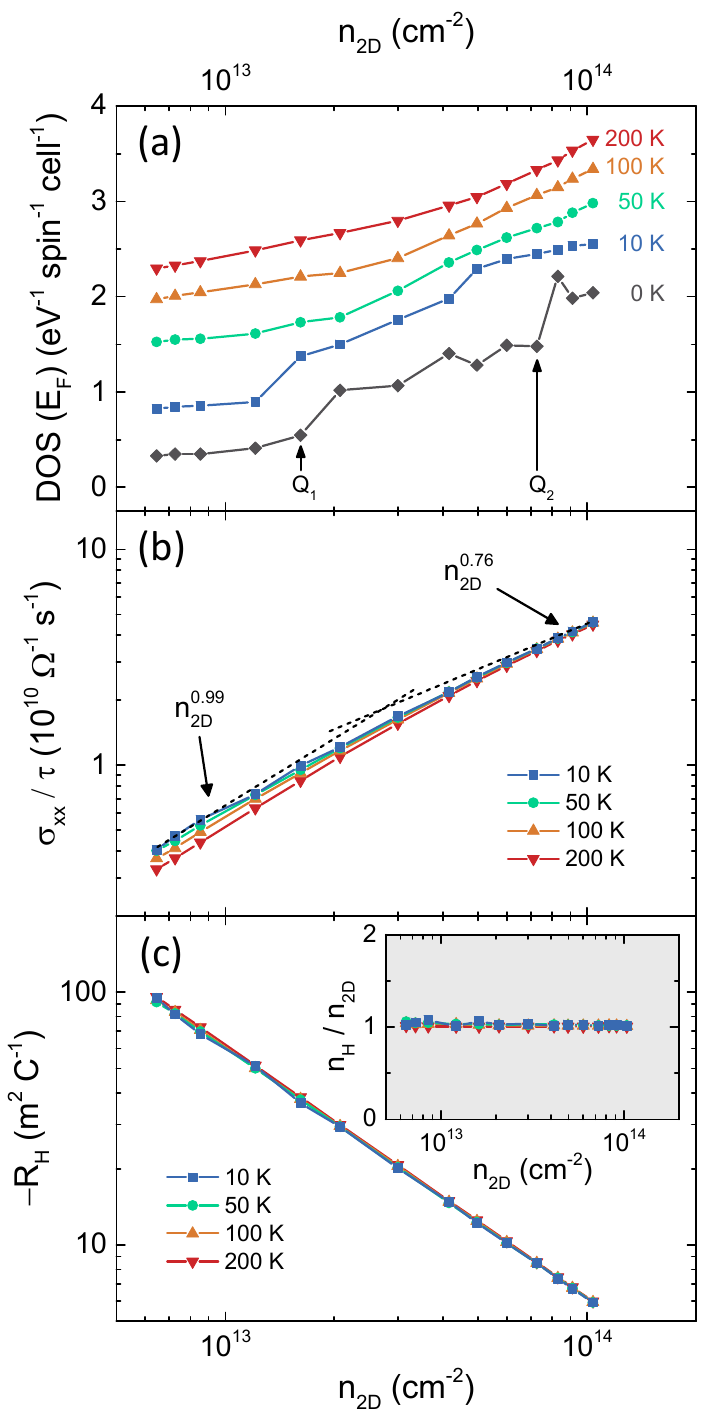}
\caption{
(a) Density of states at the Fermi level DOS$(E\ped{F})$ as a function of the doping charge density $n\ped{2D}$ for different values of $T$. Curves at finite $T$ are rigidly shifted by 0.5~eV\apex{-1}spin\apex{-1}cell\apex{-1} for clarity. 
(b) In-plane conductivity-to-lifetime ratio $\sigma_{xx}/\tau$ and 
(c) Hall coefficient $R_{xyz}$ as a function of $n\ped{2D}$ computed with BoltzTraP~\cite{BoltzTrap} for different values of $T$. The inset shows the $n\ped{2D}$-dependence of the ratio of the Hall carrier density $n\ped{H} = -1/eR_{xyz}$ and $n\ped{2D}$. Solid lines are guides to the eye. Black dashed lines in (b) highlight the power-law scaling at low and high $n\ped{2D}$.}
\label{fig:boltztrap}
\end{figure}
We first consider the effect of the band filling upon increasing electron doping on the electronic structure and transport coefficents in 4L-MoS\ped{2}. The most profound impact can be observed in the doping-dependence of the density of states at the Fermi level DOS$(E\ped{F})$, which we plot in Fig.~\ref{fig:boltztrap}a for different values of $T$. At $T = 0$~K, DOS$(E\ped{F})$ exhibits the typical staircase behavior of a multi-band 2D system, increasing in a nearly step-like fashion whenever the Fermi level crosses the bottom of a sub-band, and remaining nearly constant otherwise. By inspecting the electronic dispersion relations for increasing values of $n\ped{2D}$ shown in Fig.~\ref{fig:structure}b and in Ref.~\onlinecite{PiattiJPCM2019}, we can attribute the two sudden jumps in DOS$(E\ped{F})$ around $n\ped{2D}\approx1.5$ and $7\times10^{13}$~cm\apex{-2} to the filling of the Q\ped{1} and Q\ped{2} spin-split sub-bands respectively. At finite $T$, the 2D-like behavior of DOS$(E\ped{F})$ is quickly lost due to thermal smearing. Already at $T = 10$~K, only small ``humps" can be observed in the doping-dependence of DOS$(E\ped{F})$ in correspondence to the crossing of the Q\ped{1} and Q\ped{2} sub-bands. These humps disappear almost completely at $T = 50$~K, and at $T = 100$ and $200$~K, the doping-dependence of DOS$(E\ped{F})$ is fully smooth. This strong influence of a finite $T$ on DOS$(E\ped{F})$ can be directly attributed to the small spin-orbit splitting $\Delta_{so}$ of few meV between the sub-bands: In the K/K\apex{\prime} valleys, $\Delta_{so}$ is doping-independent and equal to about $3$~meV, whereas in the Q/Q\apex{\prime} valleys it slowly increases with doping. In particular, in the doping range before the crossing of Q\ped{2} $\Delta_{so} \lesssim 10$~meV in the Q/Q\apex{\prime} valleys~\cite{PiattiJPCM2019}.

The doping dependencies of the transport coefficients $\sigma_{xx}/\tau$ and $R_{xyz}$ are much less affected by both the band filling and $T$. Upon increasing $T$, $\sigma_{xx}/\tau$ slightly decreases in the entire doping range. For any value of $T$, $\sigma_{xx}$ smoothly increases with increasing $n\ped{2D}$ (Fig.~\ref{fig:boltztrap}b) and the only effect of band filling is to progressively reduce the power law exponent of the increase (from $\sigma_{xx}/\tau \propto n\ped{2D}^{0.99}$ for $n\ped{2D}\lesssim 1.5\times 10^{13}$~cm\apex{-2}, to $\sigma_{xx}/\tau \propto n\ped{2D}^{0.76}$ for $n\ped{2D}\gtrsim 7\times 10^{13}$~cm\apex{-2}). $R_{xyz}$, on the other hand, is found to be almost $T$-independent and smoothly decreases as $n\ped{2D}^{-1}$ in the entire doping range (Fig.~\ref{fig:boltztrap}c). Furthermore, the Hall carrier density $n\ped{H} = -1/eR_{xyz}$ is always almost identical to the doping charge $n\ped{2D}$, as shown in the inset to Fig.~\ref{fig:boltztrap}c. This is consistent with what was reported in the case of gated 1L-, 2L-, and 3L-MoS\ped{2}~\cite{BrummePRB2015, BrummePRB2016}, and is due to the good parabolicity of all sub-bands in both the K/K\apex{\prime} and Q/Q\apex{\prime} valleys and their comparable effective masses at any doping $n\ped{2D}\lesssim 2\times10^{14}$~cm\apex{-2}.

\subsection{\color{blue}Mobility and scattering lifetime}

\begin{figure}[]
\centering
\includegraphics[width=0.93\columnwidth]{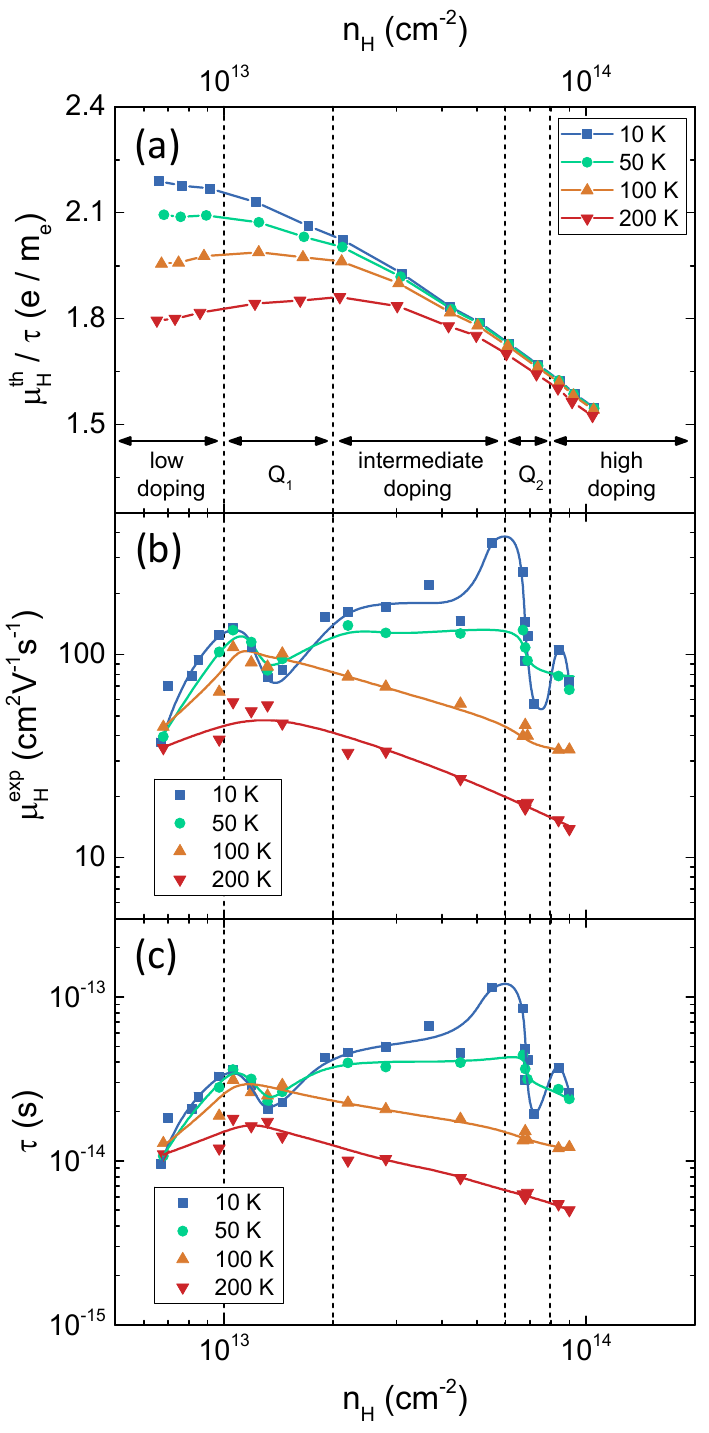}
\caption{
(a) Theoretical Hall mobility-to-lifetime ratio $\mu\ped{H}\apex{th}/\tau$, 
(b) experimental Hall mobility $\mu\ped{H}\apex{exp}$ and 
(c) scattering lifetime $\tau$ as a function of the Hall carrier density $n\ped{H}$ at different temperatures $T$. Data in (b) are directly computed from the doping-dependent conductivity values we reported in Refs.~\onlinecite{PiattiNL2018, PiattiJPCM2019}. Solid lines are guides to the eye. Vertical dashed lines highlight the different doping ranges as indicated in panel (a) and discussed in the main text.}
\label{fig:tau_determination}
\end{figure}

We now turn to the determination of the doping-dependent scattering lifetime. In Fig.~\ref{fig:tau_determination}a we show the theoretical mobility-to-scattering lifetime ratio as a function of the Hall carrier density, determined with Eq.~\ref{eq:mu_tau} from the data shown in Fig.~\ref{fig:boltztrap}. While the dependencies of $\sigma_{xx}/\tau$ and $R_{xyz}$ on $n\ped{H}$ change little upon increasing $T$, that of of $\mu\ped{H}\apex{th}/\tau$ is significantly affected instead. At low $T\leq 50$~K, $\mu\ped{H}\apex{th}/\tau$ monotonically decreases at the increase of $n\ped{H}$ and the effects of band filling are negligible. Conversely, at intermediate and high $T\geq 100$~K the $n\ped{H}$-dependence of $\mu\ped{H}\apex{th}/\tau$ becomes non-monotonic and dependent on band-filling. At low $n\ped{H}\lesssim 1.7\times10^{13}$~cm\apex{-2}, where only the K/K\apex{\prime} valleys are filled, $\mu\ped{H}\apex{th}/\tau$ increases with increasing $n\ped{H}$ and is strongly suppressed by increasing $T$. At larger $n\ped{H}\gtrsim 2\times10^{13}$~cm\apex{-2}, where also the Q/Q\apex{\prime} valleys become filled, $\mu\ped{H}\apex{th}/\tau$ decreases with increasing $n\ped{H}$ and is much less sensitive to the $T$ increase. 

On the experimental side, in Fig.~\ref{fig:tau_determination}b we show the doping-dependence of the Hall mobility of ion-gated 4L-MoS\ped{2} directly calculated from the transport data we measured in Ref.~\onlinecite{PiattiNL2018}. $\mu\ped{H}\apex{exp}$ is starkly dependent on both band filling and temperature. Before discussing them, we note that these dependencies are much stronger than those exhibited by $\mu\ped{H}\apex{th}/\tau$. As a direct consequence, the behavior of the scattering lifetime $\tau$ determined using Eq.~\ref{eq:tau_from_mu} as a function of $n\ped{H}$ and $T$ (shown in Fig.~\ref{fig:tau_determination}c) is completely dominated by that of $\mu\ped{H}\apex{exp}$. Since $\mu\ped{H}\apex{exp}$ and $\tau$ share the same dependencies, in the following we focus on discussing the behavior of $\tau$. This behavior is non-trivial, and can be separated in three main doping ranges. The first  range occurs at low doping before the crossing of Q\ped{1} ($n\ped{H}\lesssim 1\times10^{13}$~cm\apex{-2}), where $\tau$ increases with increasing $n\ped{H}$ at any $T$. The second range occurs at intermediate doping between the crossings of Q\ped{1} and Q\ped{2} ($2\lesssim n\ped{H}\lesssim 6\times10^{13}$~cm\apex{-2}), where the behavior of $\tau$ strongly depends on $T$: It increases with $n\ped{H}$ at $T=10$~K, is nearly independent of $n\ped{H}$ at $T=50$~K, and decreases with $n\ped{H}$ at $T = 100$ and $200$~K. The third range occurs at very large doping after the crossing of Q\ped{2} ($n\ped{H}\gtrsim 8\times10^{13}$~cm\apex{-2}), where $\tau$ decreases with increasing $n\ped{H}$ at any $T$. Additionally, in the narrow doping ranges corresponding to the Q\ped{1} and Q\ped{2} band crossings, $\tau$ is starkly non-monotonic below $50$~K and becomes smooth at higher $T$, mirroring the ``kinks'' observed in the doping-dependence of the conductivity~\cite{BrummePRB2016, PiattiNL2018, PiattiJPCM2019}.

\subsection{\color{blue}Scattering mechanisms}

Let's first consider the three main doping ranges away from the band crossings, where the number of bands crossing the Fermi level is constant and the electronic DOS is almost constant as well. In gated MoS\ped{2} the mobility and scattering lifetime are dominated by four main sources of scattering~\cite{YuAFM2017}: (i) acoustic phonon scattering, (ii) charged-impurity scattering, (iii) substrate-optical phonon scattering,  and (iv) charged traps. In our case, the first two mechanisms are certainly the most important, if not the only ones. Indeed, in all three doping ranges, $\tau$ decreases with increasing $T$ (except at the lowest measured value of $n\ped{H}\simeq 7\times10^{12}$~cm\apex{-2}), ruling out charged traps~\cite{YuAFM2017}. Substrate-optical phonon scattering can be also ruled out since it is weak for $T\lesssim 200$~K~\cite{YuAFM2017}, and is further suppressed in liquid-gated devices even close to room $T$~\cite{PereraACSNano2013}.

The acoustic-phonon scattering rate is expected to be doping-independent in each of the aforementioned doping ranges. Moreover, this scattering mechanism is negligible at very low $T$, and increases with $T$~\cite{YuAFM2017, PereraACSNano2013}. The scattering rate due to charged impurities, instead, is strongly doping-dependent at any $T$, since it is strongly \textit{suppressed} by the improved electrostatic screening upon increasing the carrier density~\cite{YuAFM2017}. In ion-gated devices, however, this scattering rate can also \textit{increase} upon increasing doping due to the extrinsic scattering centers introduced by the ions in the EDL, leading to a competition~\cite{GallagherNatCommun2015,  OvchinnikovNatCommun2016, PiattiApSuSc2017, LuPNAS2018, Gonnelli2DMater2017, PiattiAPL2017, PiattiEPJ2019, PiattiLTP2019, PiattiPRM2019, PiattiApSuSc2020, PiattiApSuSc2018mos2, SaitoACSNano2015}. Furthermore, in MoS\ped{2} the charged-impurity scattering rate can in general lead to a $T$-dependence of the scattering rate very similar to that due to acoustic phonon scattering~\cite{YuAFM2017}. 

\begin{figure}[]
\centering
\includegraphics[width=\mywidth]{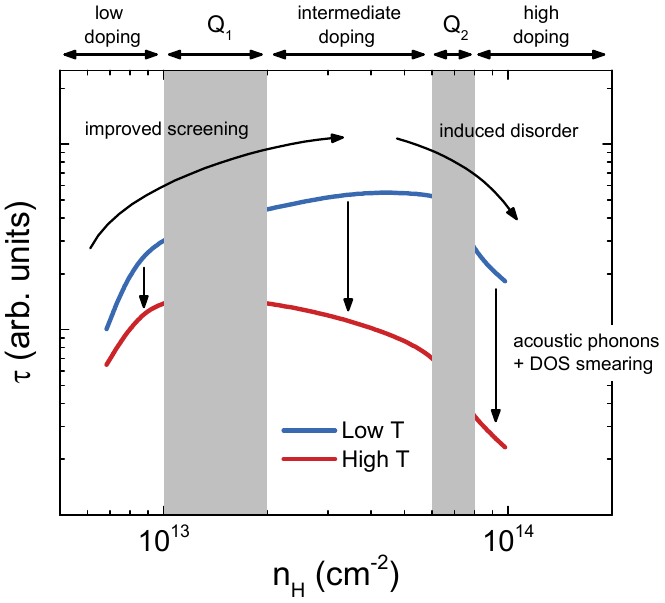}
\caption{
{\color{blue}Schematic explanation of the effects of different scattering mechanisms on 
the behavior of $\tau$ vs. $n\ped{H}$ and $T$. The doping 
ranges corresponding to the Q\ped{1} and Q\ped{2} Lifshitz transitions have been 
excluded from the analysis, so as to focus the attention on the regions 
where the number of bands crossing the Fermi level is constant. At low 
$T$, two competing mechanisms (improved screening and induced 
disorder), both ascribed to charged impurities, determine the trend of 
$\tau$, the latter being dominant at high doping. At higher $T$, 
acoustic-phonon scattering comes into play and determines a decrease in 
$\tau$, with a different magnitude in each of the three doping regions separated 
by the Lifshitz transitions. The DOS smearing further changes the shape of the 
$\tau(n\ped{H})$ curve. The same arguments apply also to the trends of 
$\mu\apex{exp}\ped{H}$.}
}
\label{fig:scattering_mechanisms}
\end{figure}

At $T = 10$~K, where the acoustic phonon scattering is negligible, the doping dependence of $\tau$ can be entirely ascribed to charged-impurity scattering. Its increase is thus due to the improved electrostatic screening; its decrease in the high-doping range, beyond the Q\ped{2} band crossing (see the last two blue points in Fig.~\ref{fig:tau_determination}c) is very likely to be due to the disorder introduced by the ions in the EDL~\cite{GallagherNatCommun2015,  OvchinnikovNatCommun2016, PiattiApSuSc2017, LuPNAS2018, Gonnelli2DMater2017, PiattiAPL2017, PiattiEPJ2019, PiattiLTP2019, PiattiPRM2019, PiattiApSuSc2020, PiattiApSuSc2018mos2, SaitoACSNano2015}. These two mechanisms are certainly acting at any $T$, but at higher temperatures the scattering from acoustic phonons suppresses $\tau$, more and more effectively as $T$ increases. The idea that phonon scattering (rather than charged-impurity scattering) is the main factor that determines the $T$ evolution of the curves is suggested by the fact that the suppression of $\tau$ is approximately uniform in each doping range. At high $T$, when the thermal smearing makes the DOS be smoothly doping dependent (see Fig.~\ref{fig:boltztrap}a), the suppression is practically uniform for any $n\ped{H}\gtrsim 1\times 10^{13}$~cm\apex{-2}. Another proof that phonon scattering dominates at high $T$ is the fact that, in the intermediate doping range, $\tau$ decreases as a function of doping, while it should increase (as it does at low $T$) if the scattering was mainly due to charged impurities. The interplay of the different scattering mechanisms is depicted schematically in Fig.~\ref{fig:scattering_mechanisms}.

\subsection{\color{blue}Intervalley scattering and Lifshitz transitions}

We now consider the two narrow doping ranges corresponding to the Q\ped{1} and Q\ped{2} band crossings where the kinks in the doping-dependent conductivity and mobility are experimentally observed in Ref.~\onlinecite{PiattiNL2018}. At a first approximation, the presence of the kinks in the conductivity and the mobility can be attributed to the strong reduction in the average Fermi velocity which occurs when the bottom of a high-energy sub-band becomes filled~\cite{BrummePRB2016, PiattiJPCM2019}. However, the reduction in the conductivity is entirely accounted for by the reduction in Fermi velocity only when the scattering lifetime is exactly inversely proportional to the density of states, $\tau \propto$~DOS$(E\ped{F})^{-1}$, for all values of doping~\cite{BrummePRB2016, PiattiJPCM2019}. Since at low $T$ in 2D systems DOS$(E\ped{F})$ follows a staircase behavior, similar kinks can be expected also in the doping-dependence of $\tau$. Indeed, as we show in Fig.~\ref{fig:tau_determination}c, these kinks do appear in the doping-dependence of $\tau$ in gated MoS\ped{2} devices, and are similarly smeared out by increasing temperature. We now investigate whether the kinks in $\tau$ can be simply explained in terms of the doping-dependence of DOS$(E\ped{F})$.

%
\begin{figure}[b]
\centering
\includegraphics[width=\mywidth]{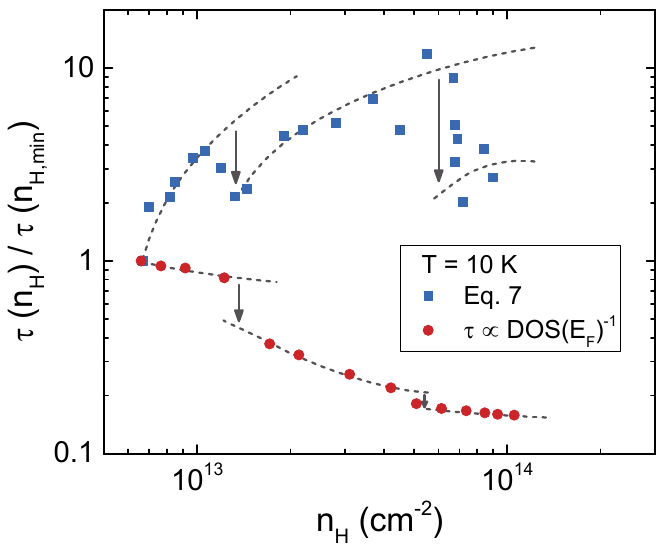}
\caption{
Scattering lifetime normalized by its value at the lowest doping, $\tau(n\ped{H})/\tau(n\ped{H,min})$, as a function of the Hall density $n\ped{H}$ at $T = 10$~K. Blue squares and red circles are obtained via Eq.~\ref{eq:tau_from_mu} and the simple approximation $\tau \propto$~DOS$(E\ped{F})^{-1}$ respectively. Black arrows highlight the drops in the scattering lifetime associated with the Q\ped{1} and Q\ped{2} band crossings. Dashed black lines are guides to the eye.
}
\label{fig:tau_discussion}
\end{figure}
To do so, we focus our attention on the data at $T=10$~K, where the kinks are most evident, and normalize the scattering lifetime by its value at the lowest Hall density, $\tau(n\ped{H})/\tau(n\ped{H,min})$ (blue squares in Fig.~\ref{fig:tau_discussion}). As highlighted by the black dashed lines, the $\tau$ at $10$~K does indeed exhibit a ``canted" staircase dependence on $n\ped{H}$, which is somewhat reminiscent of the DOS$(E\ped{F})$ computed at $T = 0$ and shown in Fig.~\ref{fig:boltztrap}a. However, when $\tau \propto$~DOS$(E\ped{F})^{-1}$ is computed from the DOS$(E\ped{F})$ at $T = 10$~K (red circles in Fig.~\ref{fig:tau_discussion}a), it becomes apparent that this simple approximation fails to reproduce most of the features of the scattering lifetime determined from the experimental mobility. $\tau \propto$~DOS$(E\ped{F})^{-1}$ is obviously unable to reproduce any \textit{increase} in $\tau$ as a function of $n\ped{H}$, since this stems from the doping-dependent charged-impurity scattering and not from the intrinsic DOS of gated MoS\ped{2}. The sudden drops in $\tau$ associated to the sub-band crossings (highlighted by the black arrows in Fig.~\ref{fig:tau_discussion}) also cannot be reproduced satisfactorily by $\tau \propto$~DOS$(E\ped{F})^{-1}$. In the case of the Q\ped{1} crossing at $T = 10$~K, the disagreement is limited: The $\tau$ determined from the experimental mobility drops by a factor $\sim 4$ upon this first Lifshitz transition, whereas $\tau \propto$~DOS$(E\ped{F})^{-1}$ estimates a smaller drop of only a factor $\sim 2$ at the same $T$. Therefore, the simple approximation correctly gauges the order-of-magnitude of the lifetime reduction, but fails in accounting for nearly half the effect observed experimentally. Most importantly, $\tau \propto$~DOS$(E\ped{F})^{-1}$ predicts that \textit{almost no drop} in $\tau$ should be observed upon crossing Q\ped{2} at $T=10$~K, in clear contrast with the $\tau$ determined from the experimental mobility. This finding is consistent with our results in Ref.~\onlinecite{PiattiJPCM2019}, where the intensity of the kink in the conductivity at Q\ped{2} was severely underestimated in a model based on $\tau \propto$~DOS$(E\ped{F})^{-1}$. Therefore, another mechanism must be responsible for the large drop in $\tau$ observed upon the second Lifshitz transition occurring due to the crossing of Q\ped{2}.

On top of increasing the DOS, filling high-energy bands can strongly alter the scattering lifetime by opening previously-forbidden \textit{interband} scattering channels, thereby strongly increasing the scattering rate~\cite{AppelPR1962}. Indeed, the kinks in the doping-dependence of the conductivity of ion-gated few-layer graphene were explicitly attributed to the activation of interband scattering by the filling of high-energy bands~\cite{YePNAS2011, GonnelliSciRep2015, PiattiApSuSc2017, Gonnelli2DMater2017}. In gated MoS\ped{2}, the evolution of the Fermi surface upon electron doping leads to the simultaneous filling of the low-energy K/K\apex{\prime} valleys and the high-energy Q/Q\apex{\prime} valleys~\cite{PiattiNL2018, PiattiJPCM2019}. Therefore, we attribute the strong reductions in the scattering lifetime to the opening of those \textit{intervalley} scattering channels that are forbidden when only the low-energy K/K\apex{\prime} valleys are populated. These include scattering channels connecting the electron pockets at Q$\leftrightarrow$Q\apex{\prime}, Q$\leftrightarrow$K, Q\apex{\prime}$\leftrightarrow$K\apex{\prime}, Q$\leftrightarrow$K\apex{\prime} and Q\apex{\prime}$\leftrightarrow$K. It is very important to note that the opening of these intervalley scattering channels has a profound influence not only on the low-$T$ scattering lifetime and mobility, but on several other key properties of gated MoS\ped{2}. 

Specifically, the availability of these intervalley scattering channels is paramount in optimizing the nesting efficiency of the Fermi surface~\cite{PickettBook}, thereby allowing to strongly enhance the electron-phonon coupling (EPC) in the system~\cite{GePRB2013, PiattiNL2018, SohierPRM2018, FuQM2017, SohierPRX2019, GarciaPRB2020, NovkoCommPhys2020}. This in turn leads to significant changes in the vibrational spectrum~\cite{GePRB2013, SohierPRX2019, SohierPRM2018}, such as the pronounced doping-dependent phonon softenings which have been observed in ion-gated MoS\ped{2} and other semiconducting TMDs by means of Raman spectroscopy~\cite{ChakrabortyPRB2012, SohierPRX2019}. In this context, the large suppression of the scattering lifetime at the crossing of the Q\ped{2} sub-band points to a dominant role of this second Lifshitz transition in the opening of intervalley scattering channels and associated strong boost to the EPC, with respect to the milder effect of the first Lifshitz transition induced by the crossing of Q\ped{1}. This is consistent with the stronger Fermi surface nesting associated with the simultaneous filling of all the available sub-bands in both the K/K\apex{\prime} and Q/Q\apex{\prime} valleys~\cite{PiattiNL2018, GarciaPRB2020}. Moreover, both the sharp increasing part of the SC dome of gated MoS\ped{2}~\cite{GePRB2013, PiattiNL2018, FuQM2017} -- which develops as a function of doping from a quantum-critical point in the same doping range where the Q\ped{2} Lifshitz transition is observed~\cite{YeScience2012, LuScience2015, FuQM2017, PiattiNL2018, ChenPRL2017} -- and the polaronic reconstruction of the Fermi sea in the K/K\apex{\prime} valleys revealed by high-resolution angle-resolved photoemission spectroscopy~\cite{KangNatMater2018, GarciaCommPhys2019}, have been explicitly attributed to the strong increase in the EPC induced by the Lifhitz transition which allows the opening of additional intervalley scattering channels. Upon further increasing the electron doping, the related Fermi surface nesting has also been predicted to become so efficient as to destabilize the 2$H$ crystal structure of pristine MoS\ped{2}~\cite{GarciaPRB2020}, thus potentially triggering the onset of a charge-density wave~\cite{RosnerPRB2014, PiattiApSuSc2018mos2} and/or a structural transition towards the 1$T$/1$T\apex{\prime}$ polytypes~\cite{PiattiApSuSc2018mos2, EdaNL2012, LinNatNano2014, LengACSNano2016, ZhuangPRB2017} and thus suppressing the SC state.

\subsection{\color{blue}Intervalley scattering and carrier localization}
\begin{figure}[]
\centering
\includegraphics[width=\mywidth]{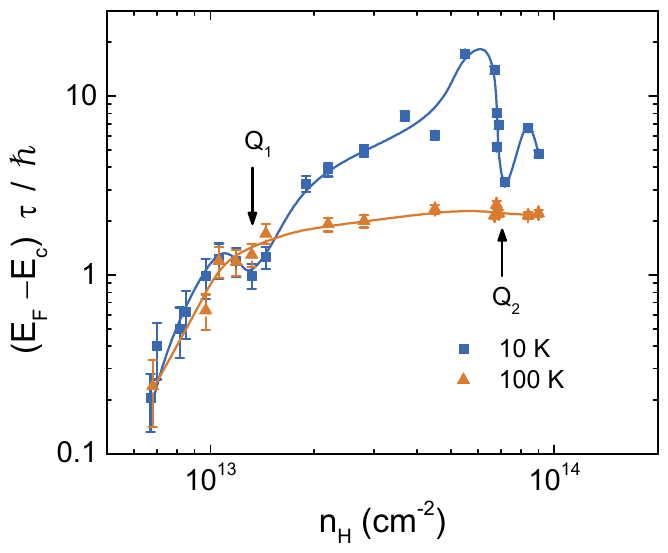}
\caption{
Ioffe-Regel parameter $(E\ped{F}-E\ped{c})\tau/\hbar$ vs. $n\ped{H}$, at $T=10$ and $100$~K. Solid lines are guides to the eye. Doping ranges where the crossings of the Q\ped{1} and Q\ped{2} sub-bands occur are highlighted.
}
\label{fig:IoffeRegel}
\end{figure}

Finally, we show that the strong suppressions of the scattering lifetime in correspondence of the Q\ped{1} and Q\ped{2} Lifshitz transitions may help in explaining another puzzling feature often observed in the two associated doping ranges in gated MoS\ped{2}. Specifically, when the kinks due to the sub-band crossings are observed in the doping-dependence of the conductivity, for those same doping levels the $T$-dependence of the resistivity often exhibits a slight upturn at very low $T$~\cite{LuScience2015, PiattiNL2018, ZheliukNatNano2019}. Ref.~\onlinecite{ZheliukNatNano2019} attributed this behavior purely to the carrier localization effect due to trap states introduced by the ions in the EDL. Our results here paint a more complex picture. While the gate-induced extra scattering centers do play a significant role in determining the scattering rate and the mobility, the largest suppressions of $\tau$ at low $T$ arise from the opening of the intervalley scattering channels (see Figs.~\ref{fig:tau_determination}c and \ref{fig:tau_discussion}). These suppressions in $\tau$ could indeed lead to carrier localization by bringing the system closer to the insulator-to-metal transition (IMT).

Following our approach in Ref.~\onlinecite{PiattiApSuSc2020}, we address this issue quantitatively by calculating the Ioffe-Regel parameter $x = (E\ped{F} - E\ped{c}) \tau / \hbar$ as a function of $n\ped{H}$ from our bandstructure calculations for 4L-MoS\ped{2} (Fig.~\ref{fig:IoffeRegel}). Here, $E\ped{F} - E\ped{c}$ is the chemical potential measured from the bottom of the conduction band $E\ped{c}$. According to the Mott-Ioffe-Regel criterion~\cite{IoffeRegel}, the Ioffe-Regel parameter characterizes the IMT in disordered systems in terms of how close the mean free path is to the lattice periodicity. When $x\gg1$ the mean free path is much larger than the lattice periodicity, leading to good metallic behavior. The opposite limit $x\ll 1$ suggests that the system is approaching the strong localization regime. The condition $x\sim1$ plays the role of a conventional crossover between the two regimes. At $T=100$~K -- where the kinks are smeared out and no resistance upturn is experimentally observed -- the Ioffe-Regel parameter increases smoothly in the whole doping range. The increase is very fast at low doping, as the gate-induced 2D electron gas (2DEG) rapidly becomes more metallic due to the filling of the K/K\apex{\prime} valleys, while it is almost constant at intermediate and high doping, likely due to the scattering lifetime being limited by electron-phonon scattering. While the 2DEG never becomes a ``good" metallic conductor ($x\geq10$) at high temperature, it is nevertheless firmly in the metallic side of the IMT as attested by its conductivity and mobility increasing with decreasing $T$. At $T=10$~K, on the other hand, the doping-dependence of the Ioffe-Regel parameter becomes non-monotonic: In the two doping ranges associated with the Q\ped{1} and Q\ped{2} band crossings, the sudden increase in the intervalley scattering rate reduces $\tau$ and brings the 2DEG back closer to the IMT. In the doping range corresponding to the crossing of Q\ped{1} the 2DEG is \textit{less} metallic at $10$~K than it is at $100$~K, whereas this inversion is not observed in the doping range corresponding to the crossing of Q\ped{2}. However, in the latter case the reduction in metallicity is comparatively much stronger and brings the 2DEG away from the ``good metal" regime reached immediately before the Lifshitz transition and back to a more localized regime. Both behaviors are consistent with a picture of incipient localization at low $T$, but are not strong enough to trigger a re-entrant IMT as in the case of gated ReS\ped{2}~\cite{OvchinnikovNatCommun2016} and WS\ped{2}~\cite{LuPNAS2018}, thus allowing for superconductivity to develop unimpeded in the system.

\section{\label{sec:conclusion}Conclusions}

In summary, we have performed \textit{ab initio} density-functional theory calculations of the bandstructure of gated and strained MoS\ped{2} nanolayers upon electron doping. We have employed the Boltzmann transport equation in the constant-relaxation-time approximation to calculate the theoretical mobility-to-scattering lifetime ratio as a function of the Hall carrier density. By combining it with the experimental data of the Hall mobility, we have determined the scattering lifetime in the system as a function of temperature and electron doping, and have discussed its behavior in terms of the major sources of charge-carrier scattering upon increasing band filling. We have shown that the scattering lifetime is strongly reduced in correspondence of the two Lifshitz transitions induced by the filling of the high-energy Q/Q\apex{\prime} valleys upon electron doping owing to the opening of additional intervalley scattering channels which become available only when both the K/K\apex{\prime} and Q/Q\apex{\prime} valleys are simultaneously occupied. We have also discussed how the opening of these intervalley scattering channels can strongly increase the electron-phonon coupling, potentially triggering the onset of the gate-induced superconducting state and of the polaronic reconstruction of the Fermi sea, as well as leading to a low-temperature incipient localization as reported in the literature.

\begin{acknowledgments}
We acknowledge funding from the MIUR PRIN-2017 program (Grant No. 2017Z8TS5B -- ``Tuning and understanding Quantum phases in 2D materials -- Quantum2D"). Computational resources were provided by hpc@polito (http://hpc.polito.it) and by CINECA, through the $'$ISCRA C$'$ project $'$HP10C8P1FI$'$.
\end{acknowledgments}

\section*{Data availability}
The data that support the findings of this study are available from the corresponding author upon reasonable request.



\begin{thebibliography}{9}
%
\bibitem{YeNatMater2010} J. T. Ye, S. Inoue, K. Kobayashi, Y. Kasahara, H. T. Yuan, H. Shimotani, and Y. Iwasa, \textit{Nat. Mater.} \textbf{9}, 125 (2010).
%
\bibitem{YeScience2012} J. T. Ye, Y. J. Zhang, R. Akashi, M. S. Bahramy, R. Arita, and Y. Iwasa, \textit{Science} \textbf{338}, 1193 (2012).
%
\bibitem{JoNanoLett2015} S. Jo, D. Costanzo, H. Berger, and A. F. Morpurgo, \textit{Nano Lett.} \textbf{15}, 1197 (2015).
%
\bibitem{ShiSciRep2015} W. Shi, J. T. Ye, Y. Zhang, R. Suzuki, M. Yoshida, J. Miyazaki, N. Inoue, Y. Saito, and Y. Iwasa, \textit{Sci. Rep.} \textbf{5}, 12534 (2015).
%
\bibitem{YuNatNano2015} Y. Yu, F. Yang, X. F. Lu, Y. J. Yan, Y.-H. Cho, L. Ma, X. Niu, S. Kim, Y.-W. Son, D. Feng, S. Li, S.-W. Cheong, X. H. Chen, and Y. Zhang, \textit{Nat. Nanotechnol.} \textbf{10}, 270 (2015).
%
\bibitem{SaitoACSNano2015} Y. Saito and Y. Iwasa, \textit{ACS Nano} \textbf{9}, 3192 (2015).
%
\bibitem{PiattiJSNM2016} E. Piatti, A. Sola, D. Daghero, G. A. Ummarino, F. Laviano, J. R. Nair, C. Gerbaldi, R. Cristiano, A. Casaburi, and R. S. Gonnelli, \textit{J. Supercond. Novel Magn.} \textbf{29}, 587-591 (2016).
%
\bibitem{LiNature2016} L. J. Li, E. C. T. O'Farrel, K. P. Loh, G. Eda, B. {\"O}zyilmaz, and A. H. Castro Neto, \textit{Nature} \textbf{529}, 185 (2016).
%
\bibitem{WangNature2016} Y. Wang, J. Xiao, H. Zhu, Y. Li, Y. Alsaid, K. Y. Fong, Y. Zhou, S. Wang, W. Shi, Y. Wang, A. Zettl, E. J. Reed, and X. Zhang, \textit{Nature} \textbf{550}, 487 (2016).
%
\bibitem{XiPRL2016} X. Xi, H. Berger, L. Forr{\'o}, J. Shan, and K. F. Mak, \textit{Phys. Rev. Lett.} \textbf{117}, 106801 (2016).
%
\bibitem{OvchinnikovNatCommun2016} D. Ovchinnikov, F. Gargiulo, A. Allain, D. J. Pasquier, D. Dumcenco, C.-H. Ho, O. V. Yazyev, and A. Kis, \textit{Nat. Commun.} \textbf{7}, 12391 (2016).
%
\bibitem{ShiogaiNatPhys2016} J. Shiogai, Y. Ito, T. Mitsuhashi, T. Nojima, and A.
Tsukazaki, \textit{Nat. Phys.} \textbf{12}, 42 (2016).
%
\bibitem{LeiPRL2016} B. Lei, J. H. Cui, Z. J. Xiang, C. Shang, N. Z. Wang, G. J. Ye, X. G. Luo, T. Wu, Z. Sun, and X. H. Chen, \textit{Phys. Rev. Lett.} \textbf{116}, 077002 (2016).
%
\bibitem{PiattiPRB2017} E. Piatti, D. Daghero, G. A. Ummarino, F. Laviano, J. R. Nair, R. Cristiano, A. Casaburi, C. Portesi, A. Sola and R. S. Gonnelli, \textit{Phys. Rev. B} \textbf{95}, 140501 (2017).
%
\bibitem{ZengNanoLett2018} J. Zeng, E. Liu, Y. Fu, Z. Chen, C. Pan, C. Wang, M. Wang, Y. Wang, K. Xu, S. Cai, X. Yan, Y. Wang, X. Liu, P. Wang, S.-J. Liang, Y. Cui, H. Y. Hwang, H. Yuan, and F. Miao, \textit{Nano Lett.} \textbf{18}, 1410 (2018).
%
\bibitem{DengNature2018} Y. Deng, Y. Yu, Y. Song, J. Zhang, N. Z. Wang, Z. Sun, Y. Yi, Y. Z. Wu, S. Wu, J. Zhu, J. Wang, X. H. Chen, and Y. Zhang, \textit{Nature} \textbf{563}, 94-99 (2018).
%
\bibitem{WangNatNano2018} Z. Wang, T. Zhang, M. Ding, B. Dong, Y. Li, M. Chen, X. Li, J. Huang, H. Wang, X. Zhao, Y. Li, D. Li, C. Jia, L. Sun, H. Guo, Y. Ye, D. Sun, Y. Chen, T. Yang, J. Zhang, S. Ono, Z. Han, and Z. Zhang, \textit{Nat. Nanotechnol.} \textbf{13}, 554-559 (2018).
%
\bibitem{PiattiPRM2019} E. Piatti, T. Hatano, D. Daghero, F. Galanti, C. Gerbaldi, S. Guastella, C. Portesi, I. Nakamura, R. Fujimoto, K. Iida, H. Ikuta, and R. S. Gonnelli, \textit{Phys. Rev. Materials} \textbf{3}, 044801 (2019).
%
\bibitem{RenNL2019} X. Ren, Y. Wang, Z. Xie, F. Xue, C. Leighton, and C. D. Frisbie, \textit{Nano Lett.} \textbf{19}, 4738-4744 (2019).
%
\bibitem{PiattiApSuSc2020} E. Piatti, A. Pasquarelli, and R. S. Gonnelli, \textit{Appl. Surf. Sci.} \textbf{528}, 146795 (2020).
%
\bibitem{KlemmBook2012} R. A. Klemm, \textit{Layered Superconductors} (Oxford University Press, Oxford, UK, 2012), Vol. 1.
%
\bibitem{KlemmPhysC2015} R. A. Klemm, \textit{Physica C} \textbf{514}, 86 (2015).
%
{\color{blue}\bibitem{WangNatNano2012} Q. H. Wang, K. Kalantar-Zadeh, A. Kis, J. N. Coleman, and M. S. Strano, \textit{Nat. Nanotechnol.} \textbf{7}, 699 (2012).
%
\bibitem{MakPRL2010} K. F. Mak, C. Lee, J. Hone, J. Shan, and T. F. Heinz. \textit{Phys. Rev. Lett.} \textbf{105}, 136805 (2010).
%
\bibitem{SplendianiNL2010} A. Splendiani, L. Sun, Y. Zhang, T. Li, J. Kim, C.-Y. Chim, G. Galli, and F. Wang, \textit{Nano Lett.} \textbf{10}, 1271-1275 (2010).
%
\bibitem{FerrariNanoscale2015} A. C. Ferrari et al., \textit{Nanoscale} \textbf{7}, 4598-4810 (2015).
%
\bibitem{MakNatPhotonics2016} K. F. Mak and J. Shan, \textit{Nat. Photonics} \textbf{10}, 216 (2016).}
%
\bibitem{ZhangNanoLett2016} R. Zhang, I.-L. Tsai, J. Chapman, E. Khestanova, J. Waters, and I. V. Grigorieva, \textit{Nano Lett.} \textbf{16}, 629 (2016).
%
\bibitem{PiattiAPL2017} E. Piatti, Q. H. Chen, and J. T. Ye, \textit{Appl. Phys. Lett.} \textbf{111}, 013106 (2017).
%
\bibitem{BiscarasNatCommun2015} J. Biscaras, Z. Chen, A. Paradisi, and A. Shukla, \textit{Nat. Commun.} \textbf{6}, 8826 (2015).
%
\bibitem{CostanzoNatNano2018} D. Costanzo, H. Zhang, B. A. Reddy, H. Berger, and A. F. Morpurgo, \textit{Nat. Nanotechnol.} \textbf{13}, 483-488 (2018).
%
\bibitem{KormanyosPRB2013} A. Korm{\'a}nyos, V. Z{\'o}lyomi, N. D. Drummond, P. Rakyta, G. Burkard, and V. I. Fal'ko, \textit{Phys. Rev. B} \textbf{88}, 045416 (2013).
%
\bibitem{YuanPRL2014} N. F. Yuan, K. F. Mak, and K. Law, \textit{Phys. Rev. Lett.} \textbf{113}, 097001 (2014).
%
\bibitem{BrummePRB2015}  T. Brumme, M. Calandra, and F. Mauri, \textit{Phys. Rev. B} \textbf{91}, 155436 (2015).
%
\bibitem{BrummePRB2016}  T. Brumme, M. Calandra, and F. Mauri, \textit{Phys. Rev. B} \textbf{93}, 081407 (2016).
%
\bibitem{KangNanoLett2017} M. Kang, B. Kim, S. H. Ryu, S. W. Jung, J. Kim, L. Moreschini, C. Jozwiak, E. Rotenberg, A. Bostwick, and K. S. Kim, \textit{Nano Lett.} \textbf{17}, 1610 (2017).
%
\bibitem{RoldanAnnPhys2014} Rold{\'a}n, J. A. Silva-Guill{\'e}n, M. P. L{\'o}pez-Sancho, F. Guinea, E. Cappelluti, and P. Ordej{\'o}n, \textit{Ann. Phys. (Berlin)} \textbf{526}, 347 (2014).
%
\bibitem{ZhaoACR2015} W. Zhao, R. M. Ribeiro, and G. Eda, \textit{Acc. Chem. Res.} \textbf{48}, 91 (2015).
%
\bibitem{LuScience2015} J. M. Lu, O. Zheliuk, I. Leermakers, N. F. Q. Yuan, U. Zeitler, K. T. Law, and J. T. Ye, \textit{Science} \textbf{350}, 1353 (2015).
%
{\color{blue}\bibitem{SaitoNatPhys2016} Y. Saito, Y. Nakamura, M. S. Bahramy, Y. Kohama, J. Ye, Y. Kasahara, Y. Nakagawa, M. Onga, M. Tokunaga, T. Nojima, Y. Yanase, and Y. Iwasa, \textit{Nat. Phys.} \textbf{12}, 144-149 (2016).}
%
\bibitem{PiattiNL2018} E. Piatti, D. De Fazio, D. Daghero, S. R. Tamalampudi, D. Yoon, A. C. Ferrari, and R. S. Gonnelli, \textit{Nano Lett.} \textbf{18}, 4821-4830 (2018).
%
\bibitem{ZhangNL2019} H. Zhang, C. Berthod, H. Berger, T. Giamarchi, and A. F. Morpurgo, \textit{Nano Lett.} \textbf{19}, 8836-8845 (2019).
%
\bibitem{BrummePRB2014} T. Brumme, M. Calandra, and F. Mauri, \textit{Phys. Rev. B} \textbf{89}, 245406 (2014).
%
\bibitem{SohierPRB2017} T. Sohier, M.Calandra, F. Mauri, \textit{Phys. Rev. B} \textbf{96}, 075448 (2017).
%
\bibitem{PiattiApSuSc2018nbn} E. Piatti, D. Romanin, R. S. Gonnelli, and D. Daghero, \textit{Appl. Surf. Sci.} \textbf{461}, 17-22 (2018).
%
\bibitem{RomaninAPSUSC2019} D. Romanin, Th. Sohier, D. Daghero, F. Mauri, R. S. Gonnelli and M. Calandra, \textit{Appl. Surf. Sci.} \textbf{496}, 143709 (2019).
%
\bibitem{RomaninApSuSc2020} D. Romanin, G. A. Ummarino, and E. Piatti, arXiv:2002.11554.
%
\bibitem{Gonnelli2DMater2017} R. S. Gonnelli, E. Piatti, A. Sola, M. Tortello, F. Dolcini, S. Galasso, J. R. Nair, C. Gerbaldi, E. Cappelluti, M. Bruna, and A. C. Ferrari, \textit{2D Mater.} \textbf{4}, 035006 (2017).
%
\bibitem{PiattiApSuSc2017} E. Piatti, S. Galasso, M. Tortello, J. R. Nair, C. Gerbaldi, M. Bruna, S. Borini, D. Daghero, and R. S. Gonnelli, \textit{Appl. Surf. Sci.} \textbf{395}, 37 (2017).
%
\bibitem{PiattiLTP2019} E. Piatti, D. Romanin, D. Daghero and R. S. Gonnelli, \textit{Low Temp. Phys.} \textbf{45(11)}, 1143-1155 (2019).
%
\bibitem{QE} P. Giannozzi et al., \textit{J. Phys. Condens. Matter} \textbf{21}, 395502 (2009).
%
\bibitem{QE_2} P. Giannozzi et al., \textit{J. Phys. Condens. Matter} \textbf{29}, 465901 (2017).
%
\bibitem{SohierPRM2018} T. Sohier, D. Campi, N. Marzari, and M. Gibertini, \textit{Phys. Rev. Materials} \textbf{2}, 114010 (2018).
%
\bibitem{BoltzTrap} G. Madsen and D. Singh, \textit{Comput. Phys. Commun.} \textbf{175}, 67 (2006).
%
\bibitem{AshcroftBook} N. Ashcroft and N. Mermin, \textit{Solid State Physics}, Science: Physics (Saunders College, Philadelphia, 1976).
%
\bibitem{PiattiJPCM2019} E. Piatti, D. Romanin, and R. S. Gonnelli, \textit{J. Phys. Condens. Matter} \textbf{31}, 114002 (2019).
%
\bibitem{BlochlPRB1994} P. E. Bl{\"o}chl, \textit{Phys. Rev. B} \textbf{50}, 17953 (1994).
%
\bibitem{PerdewPRL1996} J. P. Perdew, K. Burke, and M. Ernzerhof, \textit{Phys. Rev. Lett.} \textbf{77}, 3865 (1996).
%
\bibitem{GrimmeJPC2006} S. Grimme, \textit{J. Comput. Chem.} \textbf{27}, 1787 (2006).
%
\bibitem{MonkhorstPRB1976} H. J. Monkhorst and J. D. Pack, \textit{Phys. Rev. B} \textbf{13}, 5188 (1976).
%
\bibitem{GonnelliSciRep2016} R. S. Gonnelli, D. Daghero, M. Tortello, G. A. Ummarino, Z. Bukowski, J. Karpinski, P. G. Reuvekamp, R. K. Kremer, G. Profeta, K. Suzuki, and K. Kuroki, \textit{Sci. Rep.} \textbf{6}, 26394 (2016).
%
\bibitem{YuAFM2017} Z. Yu, Z.-Y. Ong, S. Li, J.-B. Xu, G. Zhang, Y.-W. Zhang, Y. Shi, and X. Wang, \textit{Adv. Funct. Mater.} 1604039 (2017).
%
\bibitem{PereraACSNano2013} M. M. Perera, M.-W. Lin, H.-J. Chuang, B. P. Chamlagain, C. Wang, X. Tan, M. M.-C. Cheng, D. Tom{\'a}nek, and. Z. Zhou, \textit{ACS Nano} \textbf{7}, 4449-4458 (2013).
%
\bibitem{GallagherNatCommun2015} P. Gallagher, M. Lee, T. A. Petach, S. W. Stanwyck, J.R. Williams, K. Watanabe, T. Taniguchi, and D. Goldhaber-Gordon, \textit{Nat. Commun.} \textbf{6}, 6437 (2015).
%
\bibitem{PiattiEPJ2019} E. Piatti, F. Galanti, G. Pippione, A. Pasquarelli, and R. S. Gonnelli, \textit{Eur. Phys. J. Spec. Top.} \textbf{228}, 689 (2019).
%
\bibitem{LuPNAS2018} J. Lu, O. Zheliuk, Q. Chen, I. Leermakers, N. E. Hussey, U. Zeitler, and J. Ye, \textit{Proc. Natl. Acad. Sci. USA} \textbf{115}, 3551 (2018).
%
\bibitem{PiattiApSuSc2018mos2} E. Piatti, Q. H. Chen, M. Tortello, J. T. Ye, and R. S. Gonnelli, \textit{Appl. Surf. Sci.} \textbf{461}, 269-275 (2018).
%
\bibitem{AppelPR1962} J. Appel, \textit{Phys. Rev.} \textbf{125}, 1815-1823 (1962).
%
\bibitem{YePNAS2011} J. T. Ye, M. Craciun, M. Koshino, S. Russo, S. Inoue, H. Yuan, H. Shimotani, A. F. Morpurgo, and Y. Iwasa, \textit{Proc. Natl. Acad. Sci. USA} \textbf{108}, 13002 (2011).
%
\bibitem{GonnelliSciRep2015} R. S. Gonnelli, F. Paolucci, E. Piatti, K. Sharda, A. Sola, M. Tortello, J. R. Nair, C. Gerbaldi, M. Bruna, and S. Borini, \textit{Sci. Rep.} \textbf{5}, 9554 82015).
%
\bibitem{PickettBook} W. E. Pickett. \textit{Emergent Phenomena in Correlated Matter}; Forschungszentrum J{\"u}lich GmbH and Institute for Advanced Simulations: J{\"u}lich, Germany, 2013.
%
\bibitem{GePRB2013} Y. Ge and A. Y. Liu, \textit{Phys. Rev. B} \textbf{87}, 241408 (2013).
%
\bibitem{FuQM2017} Y. Fu, E. Liu, H. Yuan, P. Tang, B. Lian, G. Xu, J. Zeng, Z. Chen, Y. Wang, W. Zhou, K. Xu, A. Gao, C. Pan, M. Wang, B. Wang, S.-C. Zhang, Y. Cui, H. Y. Hwang, and F. Miao, \textit{npj Quantum Materials} \textbf{2}, 52 (2017).
%
\bibitem{SohierPRX2019} Th. Sohier, E. Ponomarev, M. Gibertini, H. Berger, N. Marzari, N. Ubrig, and A. F. Morpurgo, \textit{Phys. Rev. X} \textbf{9}, 031019 (2019).
%
\bibitem{GarciaPRB2020} P. Garcia-Goiricelaya, J. Lafuente-Bartolome, I. G. Gurtubay, and A. Eiguren, \textit{Phys. Rev. B} \textbf{101}, 054304 (2020).
%
\bibitem{NovkoCommPhys2020} D. Novko, \textit{Commun. Phys.} \textbf{3}, 30 (2020).
%
\bibitem{GarciaCommPhys2019} P. Garcia-Goiricelaya, J. Lafuente-Bartolome, I. G. Gurtubay, and A. Eiguren, \textit{Commun. Phys.} \textbf{2}, 81 (2019).
%
\bibitem{ChakrabortyPRB2012} B. Chakraborty, A. Bera, D. V. S. Muthu, S. Bhowmick, U. V. Waghmare, and A. K. Sood, \textit{Phys. Rev. B} \textbf{85}, 161403(R) (2012).
%
\bibitem{ChenPRL2017} Q. H. Chen, J. M. Lu, L. Liang, O. Zheliuk, A. Ali, P. Sheng, and J. T. Ye, \textit{Phys. Rev. Lett.} \textbf{119}, 147002 (2017).
%
\bibitem{KangNatMater2018} M. Kang, S. W. Jung, W. J. Shin, Y. Sohn, S. H. Ryu, T. K. Kim, M. Hoesch, and K. S. Kim, \textit{Nat. Mater.} \textbf{17}, 676 (2018).
%
\bibitem{RosnerPRB2014} M. R{\"o}sner, S. Haas, and T. O. Wehling, \textit{Phys. Rev. B} \textbf{90}, 245105 (2014).
%
\bibitem{EdaNL2012} G. Eda, T. Fujita, H. Yamaguchi, D. Voiry, M. Chen,
and M. Chhowalla, \textit{Nano Lett.} \textbf{6}, 7311 (2012).
%
\bibitem{LinNatNano2014} Y.-C. Lin, D.O. Dumcenco, Y.-S. Huang, and K. Suenaga, \textit{Nat. Nanotechnol.} \textbf{9}, 391 (2014).
%
\bibitem{LengACSNano2016} K. Leng, Z. Chen, X. Zhao, W. Tang, B. Tian, C. T. Nai, W. Zhou, and K. P. Loh, \textit{ACS Nano} \textbf{10}, 9208 (2016).
%
\bibitem{ZhuangPRB2017} H. L. Zhuang, M. D. Johannes, A. K. Singh, and R. G. Hennig, \textit{Phys. Rev. B} \textbf{96}, 165305 (2017).
%
\bibitem{ZheliukNatNano2019} O. Zheliuk, J. M. Lu, Q. H. Chen, A. A. El Yumin, S. Golightly, and J. T. Ye, \textit{Nat. Nanotechnol.} \textbf{14}, 1123 (2019).
%
\bibitem{IoffeRegel} A. F. Ioffe and A. R. Regel, \textit{Prog. Semicond.} \textbf{4}, 237 (1960).
%
\end{thebibliography}
\end{document}